\documentclass{article}

\usepackage{arxiv}
\usepackage{natbib}
\usepackage[utf8]{inputenc} 
\usepackage[T1]{fontenc}    
\usepackage{hyperref}       
\usepackage{url}            
\usepackage{booktabs}       
\usepackage{amsfonts}       
\usepackage{nicefrac}       
\usepackage{microtype}      
\usepackage{lipsum}
\usepackage{amsmath}
\usepackage{graphicx}
\hypersetup{pdfborder = {0 0 0}}
\graphicspath{ {./images/} }

\title{Stock Price Prediction Using a Hybrid LSTM-GNN Model: Integrating Time-Series and Graph-Based Analysis}

\author{
 Meet Satishbhai Sonani \\
  Department of Computer Science\\
  University of Reading\\
  UK \\
   \And
 Atta Badii \\
  Department of Computer Science\\
  University of Reading\\
  UK \\
  \And
 Armin Moin \\
  Department of Computer Science\\
  University of Colorado\\
  Colorado Springs, CO, USA \\
}
\begin{document}
\maketitle
\begin{abstract}
This paper presents a novel hybrid model that integrates long-short-term memory (LSTM) networks and Graph Neural Networks (GNNs) to significantly enhance the accuracy of stock market predictions.  The LSTM component adeptly captures temporal patterns in stock price data, effectively modeling the time series dynamics of financial markets.  Concurrently, the GNN component leverages Pearson correlation and association analysis to model inter-stock relational data, capturing complex nonlinear polyadic dependencies influencing stock prices.  The model is trained and evaluated using an expanding window validation approach, enabling continuous learning from increasing amounts of data and adaptation to evolving market conditions.  Extensive experiments conducted on historical stock data demonstrate that our hybrid LSTM-GNN model achieves a mean square error (MSE) of 0.00144, representing a substantial reduction of 10.6\% compared to the MSE of the standalone LSTM model of 0.00161.  Furthermore, the hybrid model outperforms traditional and advanced benchmarks, including linear regression, convolutional neural networks (CNN), and dense networks.  These compelling results underscore the significant potential of combining temporal and relational data through a hybrid approach, offering a powerful tool for real-time trading and financial analysis.
\end{abstract}


\section{Introduction}

The stock market is a complex and dynamic system that plays a crucial role in the global economy. Stock prices are influenced by complex non-linear polyadic dependencies including economic indicators, market sentiment, geopolitical events, company-specific news, environmental events such as natural disasters, epidemics and pandemics, as well as political tensions, wars, fiscal policies, and legislation \citep{martinez2003, kullmann2002}. These factors create intricate patterns and interdependencies that make stock price movements highly unpredictable.

Accurate prediction of stock prices is of immense importance to investors, financial analysts, and policymakers. Successful predictions can lead to significant financial gains, informed decision-making, and improved economic stability. However, forecasting stock prices is inherently challenging due to the market high volatility, non-linear dynamics, and the influence of numerous interconnected factors such as economic indicators, political events, and investor sentiment \citep{fama1970, malkiel2003}. 

Traditional statistical methods for stock market prediction often fall short in capturing the intricate patterns and relationships present in financial data. Linear models, for instance, may not adequately account for non-linear dependencies and complex temporal behaviours inherent in stock price movements \citep{hiremath2010}.

In recent years, there has been a growing interest in leveraging advanced machine learning techniques to enhance predictive accuracy. Deep learning models have shown promise in modelling complex patterns in large datasets. Long Short-Term Memory (LSTM) networks, a type of Recurrent Neural Network (RNN), have been particularly effective in handling time-series data due to their ability to capture long-term dependencies and mitigate the vanishing gradient problem \citep{hochreiter1997, fischer2018}. LSTMs have been applied to stock market prediction with notable success, as they can model the sequential nature of stock prices over time.

Despite these advancements, relying solely on temporal data may not fully exploit all available information. Stock markets are influenced not only by historical prices but also by the relationships between different stocks and sectors. The performance of a stock can be affected by the performance of its competitors, suppliers, or companies within the same industry \citep{mantegna1999}. For example, during economic downturns, stocks within the same sector may exhibit correlated declines due to shared vulnerabilities. Therefore, incorporating relational data can provide additional context that enhances prediction models.

Graph Neural Networks (GNNs) have emerged as powerful tools for modelling relational data represented as graphs. By capturing the dependencies and interactions between entities, GNNs can learn complex relationships that are not readily apparent from isolated data points \citep{kipf2016, wu2021}. In the context of stock markets, GNNs can model the interconnectedness of stocks, enabling a more holistic approach to prediction. By representing stocks as nodes and their relationships as edges, GNNs can capture both direct and indirect influences among stocks.

This paper proposes a hybrid model that integrates LSTM networks and GNNs to leverage both temporal and relational data for stock market prediction. By combining the strengths of LSTMs in handling sequential data with the capabilities of GNNs in modelling relationships, the hybrid model aims to improve predictive accuracy in the volatile and complex environment of financial markets. The integration of these models addresses the limitations of traditional approaches by capturing a more comprehensive set of factors influencing stock prices.

\section{Literature Review}

\subsection{Introduction}

Accurate stock market prediction remains challenging due to market volatility, non-linearity, and sensitivity to various factors. Traditional statistical methods such as linear regression and autoregressive models often fail to capture complex dependencies and temporal dynamics in financial data \citep{gandhmal2019}. This has led to increased interest in machine learning and deep learning approaches, which handle complex data sets more effectively and adapt to market fluctuations, improving the modelling of inter-stock relationships and temporal trends \citep{senapati2018}.

Early machine learning models such as Support Vector Machines (SVM) and Artificial Neural Networks (ANN) improved on traditional methods by identifying historical patterns, but struggled with the sequential nature of financial time series. Convolutional neural networks (CNNs) excel in capturing spatial patterns but are less effective with the temporal dependencies essential for stock forecasting \citep{senapati2018}. To address these limitations, advanced models such as Long-Short-Term Memory (LSTM) networks and Graph Neural Networks (GNNs) have emerged. LSTMs model long-term dependencies in time-series data, while GNNs capture relational data among stocks. Combining these models into hybrid architectures has shown promise in analysing both temporal and relational aspects of stock data \citep{thakkar2021, ran2024}.

Explainable AI (XAI) has gained importance in financial forecasting by enhancing transparency and making AI model decisions more interpretable, which is crucial for trust and compliance in financial markets. Studies by \citet{kuiper2022} highlight that integrating XAI improves decision-making by elucidating how predictions are made.

\subsection{Data Acquisition and Challenges}

Developing reliable stock market prediction models requires accurate data acquisition, but financial data is inherently noisy, incomplete, and unpredictable, posing challenges for machine learning algorithms \citep{hadavandi2010}. Robust preprocessing techniques, including handling missing data and noise management, are essential for enhancing data quality. Optimising hyperparameters during data processing is vital, especially with large and complex datasets. Scaling and normalisation improve the accuracy of models such as LSTM by maintaining variable relationships and preventing data distortion. Detecting and managing noise in financial time series further enhances model resilience in unpredictable market conditions \citep{yeung2020}. Thus, advanced preprocessing and meticulous hyperparameter tuning are necessary for accurate stock market predictions.

\subsection{Machine Learning Approaches in Financial Forecasting}

LSTM networks have become prominent in financial time-series prediction due to their ability to capture long-term dependencies and manage the vanishing gradient problem, making them suitable for modelling the non-linear and volatile nature of stock prices \citep{fjellstrom2022}. They have been applied successfully in various financial contexts, demonstrating adaptability in recognising complex patterns and robustness in volatile markets \citep{yeung2020, zhao2023}. Comparative studies have shown that LSTM outperforms traditional models such as ARIMA in stock price forecasting, particularly with non-linear time-series data \citep{shankar2022, jarrah2023}.

GNNs are essential for analysing stock relationships by capturing dependencies between stocks, modelling the interconnectedness that traditional methods overlook. \citep{shi2024} developed a graph-based CNN-LSTM model integrating relational data with time-series analysis, achieving more accurate predictions by leveraging GNN to capture stock interconnections. Other studies have emphasised the importance of capturing relational dependencies using Graph Convolutional Networks (GCN), showing that GCN outperforms traditional time-series models by considering both temporal and relational dynamics \citep{singh2021, chen2018}. These findings illustrate that GNNs improve stock market predictions by capturing complex relationships between stocks, and when combined with models such as LSTM, they effectively handle both relational and temporal dynamics.

Hybrid models integrating LSTM networks with GNNs leverage the strengths of both methods, enabling simultaneous modelling of temporal sequences and relational data. \citet{cheng2022} demonstrated significant enhancements in predictive accuracy by combining relational data from GNN and temporal patterns from other models. \citet{shi2024} found that such hybrid models achieve more accurate predictions by capturing both temporal dynamics and inter-stock relationships. However, these models face challenges such as increased computational demands and risks of overfitting or data leakage if not properly implemented \citep{tang2021}. Careful implementation and validation are necessary to avoid these pitfalls \citep{mehtab2020}.

Dynamic modelling approaches such as rolling window and expanding window analyses are essential for adapting to evolving patterns in stock market data. Rolling window analysis trains the model on a fixed window of recent data, shifting forward with each new prediction, which is effective for short-term predictions \citep{matsunaga2019}. Expanding window analysis enlarges the training dataset by adding new data while retaining all past observations, outperforming rolling windows in capturing long-term volatility patterns \citep{feng2024}. These methods enhance the adaptability of predictive models but can introduce biases if not properly managed, necessitating a balanced approach to optimise continuous learning while maintaining historical integrity.

\subsection{Research Challenges}

Despite the advancements in stock market prediction, several outstanding research issues persist:

\begin{enumerate}
    \item \textbf{Integration of Temporal and Relational Models:} While hybrid models combining LSTM and GNN have shown potential, empirical evaluations in real-world conditions remain limited. There is a need for more extensive studies assessing their performance in volatile, real-time trading environments\citep{chen2018, shi2024}.

    \item \textbf{Robust Data Handling:} Many models overlook real-time challenges such as noisy data, missing values, and market shifts. Although preprocessing methods have been proposed \citep{bhanja2018, yeung2020}, these need further refinement to better handle the complexities of financial data and enhance model resilience.

    \item \textbf{Explainability:} The lack of transparency in LSTM and GNN models poses challenges in interpreting predictions. While Explainable AI (XAI) offers solutions \citep{kuiper2022}, its application in financial models is still limited. Integrating XAI techniques could improve trust and compliance in AI-driven financial forecasting.

    \item \textbf{Scalability and Efficiency:} Hybrid models are often computationally intensive, making real-time application difficult. Future research should focus on optimising these models for better scalability without sacrificing accuracy, possibly through algorithmic innovations or hardware acceleration.

    \item \textbf{Real-Time Adaptation:} Although expanding window analysis improves real-time predictions \citep{feng2024}, models require better strategies for continuous adaptation to new data in fast-changing markets. This includes developing methods to quickly retrain models or update predictions without extensive computational overheads.
\end{enumerate}

\section{Methodology}

\subsection{Data Collection and Preprocessing}

The dataset was obtained from Kaggle via the YFinance API, providing comprehensive historical stock data essential for time-series analysis. Ten prominent stocks representing diverse sectors were selected: Apple Inc. (AAPL), Microsoft Corporation (MSFT), Comcast Corporation (CMCSA), Costco Wholesale Corporation (COST), Qualcomm Incorporated (QCOM), Adobe Inc. (ADBE), Starbucks Corporation (SBUX), Intuit Inc. (INTU), Advanced Micro Devices (AMD), and Intel Corporation (INTC). These stocks were chosen due to their significant market capitalisation and influence, ensuring broad applicability of the findings.

\begin{figure}[ht]
    \centering
    \includegraphics[width=0.7\textwidth]{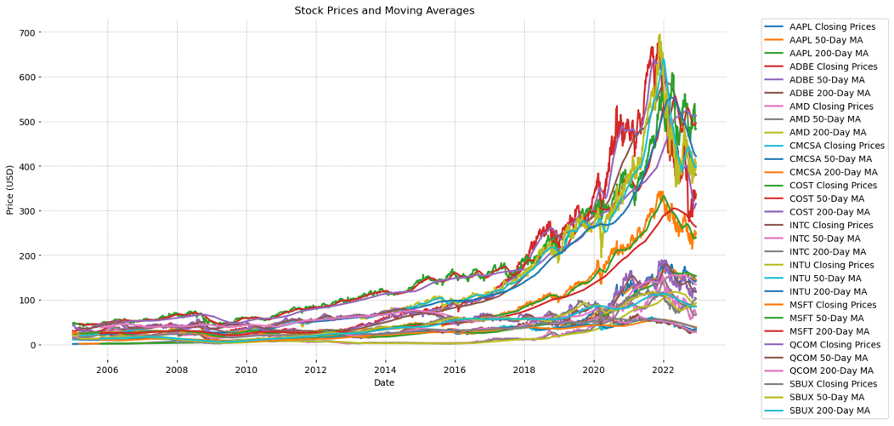}
    \caption{Normalised closing prices of a sample stock with its 50-day and 200-day moving averages.}
    \label{fig:normalised_prices}
\end{figure}

The data spanned from January 1, 2005, to December 31, 2023, encompassing various market conditions and providing a robust dataset for model training and evaluation. Features extracted included daily open, high, low, close, adjusted close prices, and trading volume, offering a detailed view of market activity. Figure~\ref{fig:normalised_prices} shows an example of normalised closing prices for a sample stock, along with its 50-day and 200-day moving averages.

\subsubsection{Feature Engineering}

To enhance predictive capabilities, feature engineering was conducted separately for the LSTM and GNN components. For the LSTM network, the primary input consisted of sequences of closing prices. Structuring the data into sequences enabled the model to capture temporal dependencies effectively. The batch size, a crucial hyperparameter, was optimised during training, with values tested at 11 and 21 to balance computational efficiency and the ability to learn long-term dependencies.

For the GNN component, a graph representing relationships between stocks was constructed. Each node represented a stock, and edges represented relationships based on Pearson correlation coefficients and association analysis. Pearson correlation captured linear relationships by calculating the correlation coefficients between the daily returns of stock pairs, while association analysis identified non-linear and complex relationships. Combining these methods provided a comprehensive representation of inter-stock relationships, crucial for modelling the interconnected nature of the stock market.

\subsubsection{Data Preprocessing}

Data preprocessing focused on scaling and structuring the data. The raw data was clean, with no missing values or significant outliers. Min-Max Scaling was applied to normalise the data. Specifically, each stock price \(x\) was transformed to a normalised value \(x'\) in the range \([0, 1]\) using

\begin{equation}
\label{eq:scaling}
x' = \frac{x - x_{\min}}{x_{\max} - x_{\min}},
\end{equation}

as shown in Equation~\ref{eq:scaling}. This normalisation was essential for the convergence of the neural network models and prevented stocks with higher absolute prices from disproportionately influencing the learning process.

Time-series batches were then created for the LSTM network, each of comprising consecutive daily closing prices.  The batch size determined the number of days of historical data instances to be ingested into the model  to enable it to make a prediction, impacting its ability to capture dependencies. Outlier detection and removal were omitted to preserve time-series integrity and reflect market realities, as extreme price movements may represent significant events rather than anomalies.

\subsection{Graphical Representation of the Stock Network}

The interdependencies between stocks were modelled using a graph \(G = (V, E)\) where each vertex \(v \in V\) represents a stock, and each edge \(e \in E\) represents a significant relationship between two stocks. Pearson correlation coefficients were calculated between the daily returns of each pair of stocks to quantify linear relationships. The daily return \(r_t\) for a stock at time \(t\) was computed as

\begin{equation}
\label{eq:daily_return}
r_t = \frac{P_t - P_{t-1}}{P_{t-1}},
\end{equation}

where \(P_t\) is the closing price at time \(t\) (see Equation~\ref{eq:daily_return}). The Pearson correlation coefficient \(\rho_{ij}\) between stocks \(i\) and \(j\) was computed as

\begin{equation}
\label{eq:pearson}
\rho_{ij} = \frac{\sum_{t=1}^{n} \left( r_{i,t} - \bar{r}_i \right) \left( r_{j,t} - \bar{r}_j \right)}{\sqrt{\sum_{t=1}^{n} \left( r_{i,t} - \bar{r}_i \right)^2} \sqrt{\sum_{t=1}^{n} \left( r_{j,t} - \bar{r}_j \right)^2}},
\end{equation}

as given in Equation~\ref{eq:pearson}. Edges were established between stocks with \(|\rho_{ij}| > 0.7\), indicating a strong linear relationship.

\begin{figure}[ht]
    \centering
    \includegraphics[width=0.4\textwidth]{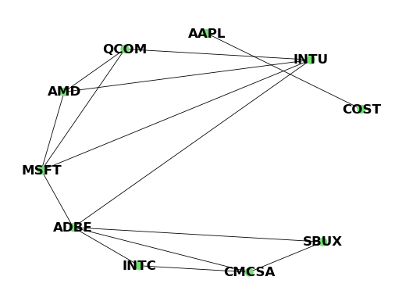}
    \caption{Graphical representation of the stock network.}
    \label{fig:stock_graph}
\end{figure}

To capture non-linear relationships, association analysis was employed via the Apriori algorithm.  In addition to support (the fraction of days over which two stocks moved together) and confidence (the probability of the movement of one stock responsive to the price fluctuations of other stocks), lift was used to gauge whether the co-movement was stronger than random chance. A lift threshold of 1.7 was set, meaning that two stocks appeared 70\% more frequently together than if they were independent. Only rules exceeding this threshold (and meeting minimum support and confidence criteria) contributed additional edges in the final undirected graph. Figure~\ref{fig:stock_graph} illustrates the constructed stock network.

\subsection{LSTM Component}

The LSTM network captured temporal dependencies in stock prices by processing sequences of historical data. The architecture included an input layer that accepted normalised closing price sequences over optimised time windows. Multiple LSTM layers were stacked to learn complex temporal patterns, utilising gating mechanisms (forget, input, and output gates) along with cell state updates and hidden states. A dense layer subsequently transformed the LSTM outputs into feature vectors for integration with the GNN component.

Hyperparameters such as learning rate, batch size, and the number of epochs were optimised through experimentation. Learning rates of 0.001, 0.005, and 0.01 were tested, with batch sizes of 11 and 21 evaluated for optimal sequence length. The number of epochs varied between 10 and 50, employing early stopping to prevent overfitting. The Adam optimiser was used for efficient training and Mean Squared Error (MSE) served as the loss function.

\subsection{GNN Component}

The GNN component modelled relational dependencies among stocks based on the constructed graph. The architecture comprised an input layer that received the stock graph and initial node features (including stock-specific attributes). Two graph convolutional layers aggregated information from neighbouring nodes using degree-normalised message passing, ensuring balanced influence among nodes. The output layer generated updated node embeddings that incorporated both the features of the stock and the aggregated neighbour information.

Edges were weighted according to the strength of relationships derived from Pearson correlation and association analysis. ReLU activation functions introduced non-linearity after each graph convolutional layer, and dropout techniques were applied for regularisation. Dropout rates were optimised during training to prevent overfitting.

\subsection{Hybrid Model Integration}

The hybrid model integrated outputs from the LSTM and GNN components to leverage both temporal and relational information. Temporal embeddings from the LSTM captured historical price patterns, while relational embeddings from the GNN encapsulated inter-stock relationships. These embeddings were concatenated to form a unified feature vector, which was then passed through additional dense layers to learn complex interactions. A final dense layer with a linear activation function produced the predicted closing price. Hidden layers utilised ReLU activations to capture non-linear relationships, and the output layer employed a linear activation suitable for regression tasks. The model was trained using the MSE loss function with the Adam optimiser.

\subsection{Training Strategy}

\begin{figure}[ht]
    \centering
    \includegraphics[width=0.5\textwidth]{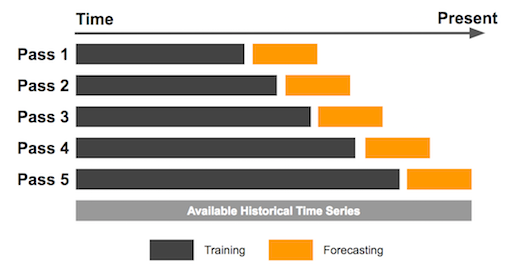}
    \caption{Expanding window approach visualisation \citep{worsnup2022}}
    \label{fig:expanding_window}
\end{figure}

To evaluate the performance of the model in a manner that reflects real-world trading scenarios, an expanding window validation strategy was implemented. Baseline models (including linear models, CNNs, dense neural networks, and standalone LSTM models) were first trained to provide a comparative benchmark. The overall training period was set to two years, with 50 days reserved for testing. In this setup, the model was tested on one day at a time; after each test, the data from the corresponding day was added to the training set, effectively expanding the training window. This iterative process, illustrated in Figure~\ref{fig:expanding_window}, ensured that the model consistently incorporated new information from the most recent market conditions, reflecting a dynamic and adaptive learning process. The model was retrained at each step with the updated dataset, enabling it to learn from evolving market trends and continuously improve its predictions. This strategy prevented data leakage by ensuring that only historical data was used for training and enhanced the robustness of the model.

\subsection{Training Parameters}

\begin{itemize}
    \item \textbf{Early Stopping:} Implemented based on validation loss to prevent overfitting, with a patience parameter set to halt training if no improvement was observed over several epochs.
    \item \textbf{Number of Epochs:} Varied between 10 and 50, with the optimal number determined through experimentation.
    \item \textbf{Batch Size:} Maintained consistency with the batch sizes used during model input preparation.
\end{itemize}

\section{Experiments and Results}

\subsection{Experiment Setup}

The experiments were conducted on a high-performance computing platform equipped with an NVIDIA GTX 1080 GPU (8 GB VRAM), 16 GB of RAM, and a multi-core Intel i7 processor. This hardware configuration was essential for handling the computational demands of the hybrid LSTM-GNN model. The software environment included Windows OS and Python 3.8. Key libraries used were PyTorch and PyTorch Geometric for model implementation, NumPy and Pandas for data manipulation, scikit-learn for baseline models, NetworkX for graph construction, and Matplotlib for data visualisation. This setup ensured efficient management of both time-series and graph-based data.

Hyperparameter tuning was a critical aspect of the experiment. A grid search was conducted to identify the optimal hyperparameters. The learning rates tested were 0.001, 0.005, and 0.01, with 0.005 providing the best balance between convergence speed and stability. Batch sizes of 11 and 21 were evaluated, with a batch size of 11 selected for its effectiveness in capturing temporal dependencies while maintaining computational efficiency. The number of epochs ranged from 10 to 50, with optimal performance achieved between 40 and 50 epochs.

Early stopping criteria were applied using a patience parameter and a minimum delta to halt training when improvements became negligible, thus optimising model performance and preventing overfitting. The Adam optimiser was employed for its adaptive learning rate capabilities. ReLU activation functions were used in hidden layers to introduce non-linearity, while a linear activation function was applied in the output layer—suitable for the regression task of stock price prediction. Dropout layers with a rate of 0.5 were incorporated in both the LSTM and GNN components to mitigate overfitting.

The Mean Squared Error (MSE) was chosen as the primary evaluation metric, defined as

\begin{equation}
\label{eq:mse}
\text{MSE} = \frac{1}{n}\sum_{i=1}^{n} \left(\hat{y}_i - y_i\right)^2,
\end{equation}

where \( n \) is the number of predictions, \(\hat{y}_i\) is the predicted stock price for the \(i\)th data point, and \(y_i\) is the actual stock price for the \(i\)th data point (see Equation~\ref{eq:mse}). MSE was selected for its sensitivity to large errors—a critical aspect in financial applications where significant deviations can result in substantial financial losses.

\subsection{Performance Analysis}

The hybrid LSTM-GNN model demonstrated strong performance in stock price prediction by effectively integrating time-series data (captured by the LSTM component) and relational data (modelled by the GNN). By accounting for both the temporal patterns of individual stocks and the interdependencies among different stocks, the model achieved enhanced predictive accuracy. Consistently low MSE values were observed across most test days.

\begin{figure}[ht]
    \centering
    \includegraphics[width=0.99\textwidth]{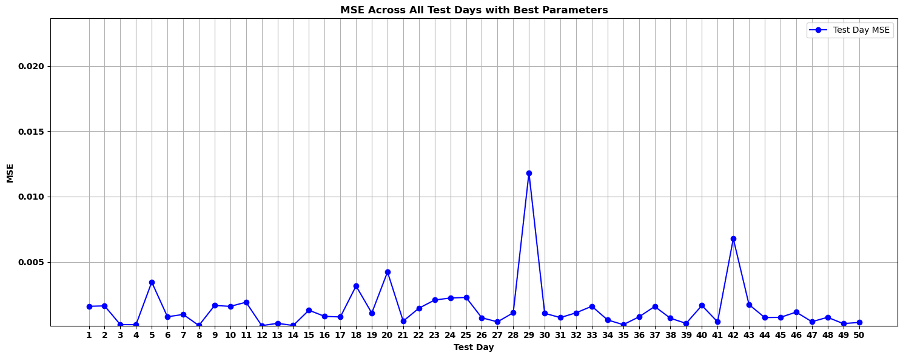}
    \caption{MSE values across all test days using the best parameter configuration.}
    \label{fig:mse_test_days}
\end{figure}

Figure~\ref{fig:mse_test_days} presents the MSE values across all test days using the best parameter configuration. Notably, two significant spikes in MSE were observed on November 10, 2022, and November 30, 2022. These spikes, as depicted in Figure~\ref{fig:mse_test_days}, can be attributed to abrupt market volatility triggered by external factors such as major financial news, earnings reports, or geopolitical events.

\subsection{Impact of Expanding Window Validation}

The expanding window validation strategy played a crucial role in adapting the model to changing market conditions. By continually updating the training set with new data, the model maintained dynamic, real-time applicability. This method enabled the model to balance between retaining historical trends and incorporating recent market dynamics, ensuring that predictions remained reflective of current market behaviour. The expanding window approach not only enhanced backtesting accuracy but also provides a clear path toward real-world deployment, where constant data updates are necessary. Over time, exposing the model to a larger and more diverse dataset also helped mitigate overfitting and improved generalisation.

\subsection{Hyperparameter Tuning}

Hyperparameter tuning was conducted using a grid search to identify the optimal combination of learning rates, batch sizes, and epochs. The final configuration that yielded the best performance consisted of a learning rate of 0.005, 40 epochs, and a batch size of 11. This setup produced the lowest MSE, indicating effective convergence while minimising prediction errors. Early stopping was applied during training to halt the process if no improvement in validation loss was observed for five consecutive epochs, thereby optimising both training time and model performance.

\subsection{Comparison with Baseline Models}

\begin{figure}[ht]
    \centering
    \includegraphics[width=0.8\textwidth]{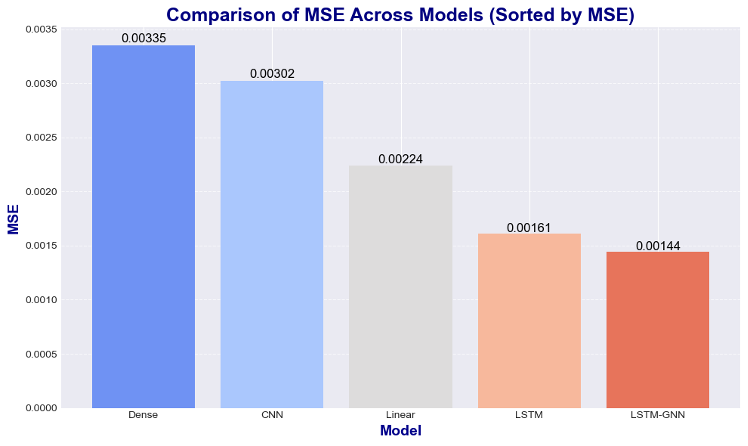}
    \caption{Comparison of MSE values across different models.}
    \label{fig:model_comparison}
\end{figure}

The hybrid LSTM-GNN model was evaluated against several baseline models, including Linear Regression, Convolutional Neural Networks (CNN), Dense Neural Networks (DNN), and a standalone LSTM. The hybrid model outperformed all baselines in terms of MSE. \\\\
Figure~\ref{fig:model_comparison} compares the MSE values across the different models. The hybrid LSTM-GNN achieved the lowest MSE of 0.00144. In comparison, the Linear Regression model recorded an MSE of 0.00224, while the standalone LSTM achieved an MSE of 0.00161. The CNN and DNN models underperformed, with MSE values of 0.00302 and 0.00335, respectively, due to their limited ability to capture temporal dependencies and complex inter-stock relationships.

\begin{figure}[ht]
    \centering
    \includegraphics[width=0.7\textwidth]{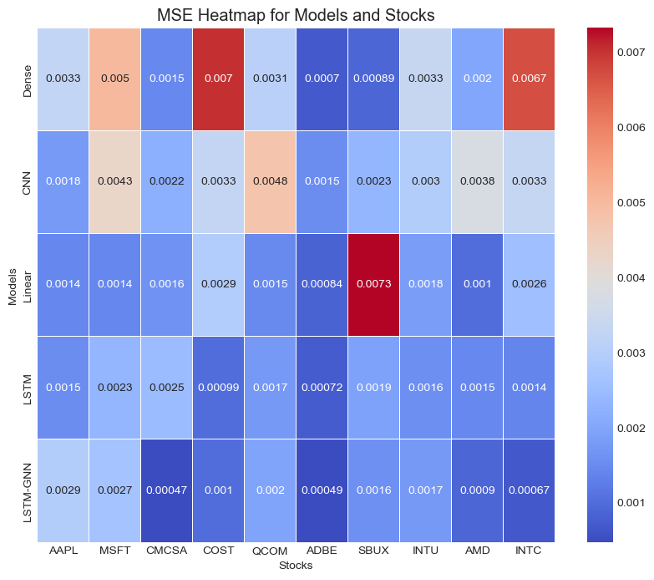}
    \caption{Heatmap of MSE values for different models across individual stocks.}
    \label{fig:stock_heatmap}
\end{figure}

Further analysis examined model performance on individual stocks. Figure~\ref{fig:stock_heatmap} presents a heatmap of the MSE values for each model across the ten stocks analysed. The hybrid model consistently achieved lower MSE values across all stocks, demonstrating its robustness and generalisability. Notably, the hybrid model performed exceptionally well for stocks such as CMCSA, AMD, and INTC, while the CNN and DNN models exhibited higher MSE values, particularly for more volatile stocks.

\subsection{Comparative Study}

The comparative analysis demonstrates the clear advantages of the hybrid LSTM-GNN model over baseline models in predictive accuracy and robustness. While the standalone LSTM captured temporal dependencies effectively, it lacked the ability to model inter-stock relationships significantly influencing market behaviour. Incorporating the GNN component enabled the hybrid model to utilise relational data, capturing complex interactions between stocks and enhancing predictions.

Compared to the standalone LSTM, the hybrid model achieved a 10.6\% reduction in average MSE (0.00144 vs 0.00161). This improvement highlights the value of incorporating relational data through the GNN component, which captures both linear and non-linear relationships between stocks. The added contextual information from the GNN enabled the model to leverage insights from stock correlations and broader market trends, improving overall predictive performance.

The DNN and CNN models, although capable of modelling non-linear relationships, underperformed due to their inability to capture the sequential nature of stock price movements. Stock prices are inherently temporal, and models that do not account for this structure often fail to identify critical patterns. Consequently, the LSTM-based models outperformed the DNN and CNN, reinforcing the importance of temporal modelling in financial forecasting.

Linear Regression, as a simple baseline model, showed limitations in capturing non-linear relationships and temporal dependencies, resulting in an MSE of 0.00224. Although its MSE was lower than that of the CNN and DNN models, it was less effective than the LSTM-based models due to its inability to model the intricate dynamics of stock markets. These limitations were especially evident in more volatile stocks, where Linear Regression struggled with complex market movements.

A key factor driving the hybrid model superior performance was the expanding window training approach. This method progressively increased the training dataset by incorporating new data as it became available, enabling the model to remain up-to-date with recent market trends. Retraining with the most current data enabled the hybrid LSTM-GNN to continuously adapt to changes in market behaviour.

The expanding window training approach offers several key advantages. Firstly, it enhances adaptability by enabling the model to learn from recent patterns and market anomalies, which improves predictive accuracy in a dynamic environment. Additionally, it mitigates concept drift by ensuring the training data reflects current market conditions, thereby maintaining the relevance of the model predictions. This approach also improves generalisation by exposing the model to a more diverse dataset, reducing the risk of overfitting and enhancing its ability to perform well on unseen data.

In the hybrid model, the LSTM component benefited from extended data sequences, capturing long-term dependencies and temporal dynamics more effectively. Simultaneously, the GNN component adapted to evolving inter-stock relationships, such as changing correlations or emerging market sectors. This dual advantage made the hybrid model exceptionally effective in handling the complexities of stock prediction.

Despite its strong performance, the hybrid model presented certain limitations. The integration of LSTM and GNN components, combined with expanding window training, resulted in higher computational demands. The complexity required more time and resources to train compared to simpler baseline models, which could be challenging in real-time trading environments requiring rapid predictions. Frequent retraining as new data became available further increased the computational burden. The model was also sensitive to hyperparameter configurations, necessitating careful tuning to achieve optimal results.

The significant improvement in predictive accuracy offered by the hybrid model has important implications for financial decision-making. More accurate stock price predictions enable traders and investors to make better-informed decisions, potentially leading to higher returns and reduced risk. The model ability to capture both temporal and relational patterns makes it a powerful tool for navigating uncertainties in stock markets and adapting to shifting market conditions.

\section{Discussion}

\subsection{Interpretation of Results}

The hybrid LSTM-GNN model significantly outperformed baseline models in stock price prediction due to its ability to capture both temporal dynamics and relational dependencies. The LSTM component modelled sequential patterns and long-term trends in stock prices, essential in financial time-series data where past events influence future movements. The GNN component captured complex inter-stock relationships by constructing a graph based on Pearson correlation and association analysis, accounting for both linear and non-linear dependencies.

By integrating temporal and relational embeddings, the hybrid model leveraged the strengths of both approaches, resulting in lower Mean Squared Error (MSE) compared to the standalone LSTM. The expanding window training approach enhanced adaptability by continuously incorporating new data, ensuring the model remained attuned to recent market conditions—a critical factor in the dynamic financial environment. The robust performance of the model across various stocks, including those which are highly volatile such as AMD, indicates its effectiveness in capturing both stable patterns and sudden market shifts.

\subsection{Limitations}

Despite its advantages, the hybrid model has limitations. The increased computational complexity from integrating LSTM and GNN components demands significant processing power and memory, which may not be readily available to all practitioners. The expanding window approach, while improving adaptability, complicates validation since it lacks a separate validation set, increasing the risk of overfitting without proper feedback during training.

The model performance is sensitive to hyperparameter tuning, requiring extensive experimentation that can be resource-intensive. Data limitations, such as missing values or anomalies, can degrade the model effectiveness. Additionally, assuming that past relationships persist into the future may not hold during unprecedented market events or structural economic changes, reducing the model predictive accuracy.

Frequent retraining due to the expanding window method increases computational load, which can be impractical in real-time applications where swift predictions are essential. This could limit the model applicability in high-frequency trading environments that demand rapid decision-making.

\subsection{Implications for Practice}

The enhanced predictive accuracy of the hybrid model holds significant implications for real-time trading and financial analysis. It can aid traders and investors in making informed decisions, improving risk management and potentially increasing returns. By accurately forecasting stock prices, the model supports strategies such as algorithmic trading, portfolio optimisation, and risk assessment.

Implementing the model in real-time applications requires addressing computational efficiency. Techniques such as incremental learning or online updating could reduce the need for full retraining, decreasing computational overhead and improving response times. Extending the model architecture to other financial instruments - 
 such as commodities, cryptocurrencies, or foreign exchange rates - could validate its adaptability, as these markets also exhibit temporal and relational dynamics.

Incorporating additional data sources, such as macroeconomic indicators, news sentiment, or social media trends, could further enhance predictive capabilities. The graph-based approach offers flexibility in modelling various relationships, including supply chains, industry classifications, or geographical linkages, providing deeper insights into the financial ecosystem.

For institutional investors, the model serves as a sophisticated tool for analytics and forecasting. It can assist in stress-testing portfolios under different market scenarios by simulating how shocks propagate through the network, which is valuable for risk management and regulatory compliance.

\section{Conclusion}

This study introduced a hybrid Long Short-Term Memory (LSTM) and Graph Neural Network (GNN) model for stock price prediction, integrating temporal and relational data to enhance predictive accuracy. The key findings demonstrate that the hybrid model significantly outperforms traditional models such as linear regression, convolutional neural networks, dense neural networks, and standalone LSTM models. The hybrid model achieved a notable reduction in Mean Squared Error (MSE), approximately 10.6\% lower than the standalone LSTM, highlighting the effectiveness of incorporating inter-stock relationships through the GNN component.

The primary contribution of this research is the demonstration of how integrating temporal dependencies with relational information can lead to more accurate stock price predictions. By utilising the expanding window training approach, the model remained adaptive to evolving market conditions, further enhancing its performance. This approach addresses the non-stationary nature of financial markets, enabling the model to capture both long-term trends and recent market shifts.

For future research, several avenues can be explored to build upon the findings of this study. Incorporating additional data sources such as macroeconomic indicators, news sentiment analysis, or social media trends could provide a more comprehensive understanding of the factors influencing stock prices. Refining the model architecture to improve computational efficiency would make it more suitable for real-time trading applications. Additionally, extending the model to predict other financial instruments, such as commodities, cryptocurrencies, or foreign exchange rates, could validate its applicability across different markets.

In conclusion, the hybrid LSTM-GNN model presents a significant advancement in stock price prediction by effectively capturing the complexities of financial data. This research contributes to the field by demonstrating the value of combining temporal and relational modelling, offering a promising direction for future developments in financial forecasting.

\bibliographystyle{agsm}



\end{document}